\documentclass{pjab}
\usepackage[T1]{fontenc}
\usepackage{lmodern}
\usepackage{textcomp}
\usepackage{amsmath}
\usepackage[dvipdfmx]{graphicx}
\usepackage{color}
\usepackage{array}
\usepackage[superscript]{cite}
\usepackage{hyperref}

\begin{document}
\pjabcategory{Review}
\title[Diffuse supernova neutrino background]
      {Diffuse neutrino background from past core-collapse supernovae}
\authorlist{%
 \Cauthorentry{Shin'ichiro Ando}{labelA,labelB}
 \authorentry{Nick Ekanger}{labelC}
 \authorentry{Shunsaku Horiuchi}{labelC,labelB}
 \authorentry{Yusuke Koshio}{labelD,labelB}
}
\affiliate[labelA]{GRAPPA Institute, University of Amsterdam, 1098 XH Amsterdam, The Netherlands}
\affiliate[labelB]{Kavli Institute for the Physics and Mathematics of the Universe (WPI), University of Tokyo, Chiba 277-8583, Japan}
\affiliate[labelC]{Center for Neutrino Physics, Department of Physics, Virginia Tech, Blacksburg, Virginia 24061, USA}
\affiliate[labelD]{Department of Physics, Okayama University, Okayama, Okayama 700-8530, Japan}
\Correspondence{Shin'ichiro Ando (s.ando@uva.nl)}
\abstract{Core-collapse supernovae are among the most powerful explosions in the universe, emitting thermal neutrinos that carry away the majority of the gravitational binding energy released. These neutrinos create a diffuse supernova neutrino background (DSNB), one of the largest energy budgets among all radiation backgrounds. Detecting the DSNB is a crucial goal of modern high-energy astrophysics and particle physics, providing valuable insights in both core-collapse modeling, neutrino physics, and cosmic supernova rate history. In this review, we discuss the key ingredients of DSNB calculation and what we can learn from future detections, including black-hole formation and non-standard neutrino interactions. Additionally, we provide an overview of the latest updates in neutrino experiments, which could lead to the detection of the DSNB in the next decade. With the promise of this breakthrough discovery on the horizon, the study of DSNB holds enormous potential for advancing our understanding of the Universe.}
\maketitle

\section{Introduction}

It has been 35 years since a supernova exploded in the neighbourhood of our Milky-Way galaxy. This supernova, SN~1987A, which happened in the Large Magellanic Cloud 50~kpc away from the Earth, gave the first detected neutrino signal from beyond the solar system. In total, 11, 8, and 5 neutrino events were detected by Kamiokande-II\cite{Kamiokande-II:1987idp}, IMB\cite{Bionta:1987qt}, and Baksan\cite{Alekseev:1988gp} detectors, respectively, enabling intensive discussions on the core-collapse mechanisms and neutrino physics.
The next Galactic core-collapse supernova will be a spectacular event for not just neutrino telescopes and various traditional astronomical observatories but also gravitational wave detectors; it will quite possibly be the first multi-messenger transient where all three messengers are detected. 
If this happens at the Galactic center, this will yield thousands of neutrino events at the current generation of water Cherenkov detectors such as Super-Kamiokande.
Other detectors will also detect hundreds of neutrino events, covering a wide energy range and different flavor species altogether.
For example IceCube, a Gton-scale neutrino telescope at the south pole, and its planned successor, IceCube-Gen2, are sensitive to the time evolution of neutrino luminosity very precisely, even though they do not have sensitivity to measure the energy of each neutrino.
However, since the supernova rate is estimated to be a few per century\cite{Rozwadowska:2020nab}, the probability of having such an event over the coming decade is unfortunately small, around 10--30\%.

Even in an unlucky situation where we do not have a Galactic supernova in the near future, detector sensitivities have improved substantially such that we are in a position to detect neutrinos from supernovae at cosmological distances. 
Within the Hubble volume, the rate of core-collapse supernovae is on the order of one per second, which compensates the reduction of flux from each supernova.
Furthermore, the detection of this Diffuse Supernova Neutrino Background (DSNB) is just around the corner, especially after the gadolinium upgrade of the Super-Kamiokande detector.
By measuring the spectrum of the DSNB, one can study various important physics of both the core-collapse supernovae and neutrinos themselves.
For example, the DSNB spectrum will contain information on the cosmic history of supernova rate evolution, neutrino generation mechanism, and the fraction of black holes that form as a result of the core collapse (in which case there may not be a supernova associated to it).
It also provides an ideal setup for the longest baseline low-energy neutrino experiments to test various new interactions in the neutrino sector with unprecedented sensitivity.

All these physics ingredients are combined to predict the flux of the DSNB spectrum. Since there is no time or directional information available for measurement\footnote{Supernovae are stochastic events. However, since their frequency of $\sim$1~Hz is way higher than frequency of the DSNB detection, the DSNB events will look isotropic and diffuse.}, only the energy distribution (i.e., spectrum) is the observable for the DSNB. 
Therefore, there is a high degree of degeneracy for these model ingredients. 
To enable detailed interpretations, one must have sufficient statistics and also the most accurate theoretical models.
The former can be achieved with large-volume neutrino telescopes. 
A prime example is the Super-Kamiokande experiment based in Japan, which recently restarted taking neutrino data with Gadolinium to substantially reduce the backgrounds.
The latter relies on progress in supernova physics mainly driven by numerical simulations of the core-collapse process and neutrino physics that progressed substantially over the previous decades, especially for the oscillation parameters, as well as astronomical observables of supernova and associated quantities. 

In this article, we review the studies of the DSNB by going over the latest progress in the field. Predictions have improved steadily over the years~\cite{Krauss:1983zn,Dar:1984aj,Totani:1995rg,Totani:1995dw,Malaney:1996ar,Hartmann:1997qe,Kaplinghat:1999xi,Ando:2002ky,Ando:2002zj,Fukugita:2002qw,Strigari:2003ig,Iocco:2004wd,Strigari:2005hu,Lunardini:2005jf,Daigne:2005xi,Yuksel:2005ae,Horiuchi:2008jz,Lunardini:2009ya,Lien:2010yb,Keehn:2010pn,Vissani:2011kx,Lunardini:2012ne,Nakazato:2013maa,Mathews:2014qba,Yuksel:2012zy,Nakazato:2015rya,Hidaka:2016zei,Priya:2017bmm,Horiuchi:2017qja,Moller:2018kpn,Riya:2020wpw,Kresse:2020nto,Ekanger:2022neg,Ziegler:2022ivq,Ashida:2022nnv,Ashida:2023heb,Anandagoda:2023sbg} 
and have been the focus in several past reviews~\cite{Ando:2004hc, Beacom:2010kk, Lunardini:2010ab} as well as reviews focusing on particular aspects of the DSNB.\cite{Mathews:2019klh}
We will also cover the state-of-the-art of the Super-Kamiokande detector and its sensitivity to study supernova and neutrino physics.

\section{Formalism}

\subsection{Brightness of the extragalactic neutrino sky}

Before deriving the rigorous formula with which one can compute the DSNB flux, we begin with a simple order-of-magnitude estimate.
Each core-collapse supernova releases the 99\% of its gravitational binding energy on the order of $E_b = 3GM_{\rm NS}^2/(5R_{\rm NS}) = 3\times 10^{53}$~erg in the form of thermal neutrinos.
Here $M_{\rm NS} = 1.4 M_\odot$ and $R_{\rm NS} = 10$~km are the mass and radius of a newly born neutron star, respectively.
The supernova rate in the Milky Way is estimated to be $\sim 0.02$~yr$^{-1}$.
Combined with the local number density of galaxies that is estimated to be $0.01$~Mpc$^{-3}$, the local supernova rate density is $\rho_{\rm SN} = 2\times 10^{-4}$~yr$^{-1}$~Mpc$^{-3}$.
The Universe has kept injecting supernova neutrinos since the beginning of star formation, and the time scale can be estimated as the Hubble time $t_H \equiv H_0^{-1}$, where $H_0$ is the Hubble constant.
By multiplying the three we obtain the energy density of the DSNB as $\epsilon_{\rm DSNB} = E_b \rho_{\rm SN} t_H \approx 3\times 10^{-14}$~erg~cm$^{-3}$.
This is likely a conservative estimate because the rate of supernovae is known to be larger in the past, by up to an order of magnitude at the redshifts $z = 1$--2.

The brightest radiation component in the Universe is the cosmic microwave background (CMB) --- the radiation from the Big Bang --- whose energy density is $\epsilon_{\rm CMB} = 4\times 10^{-13}$~erg~cm$^{-3}$. The second brightest component after the CMB is the extragalactic background light (EBL; i.e., emissions from stars and those reprocessed via dust absorption and re-emission) covering the infrared, optical, and ultraviolet wavebands and its energy density is $\epsilon_{\rm EBL} = (2$--$3) \times 10^{-14}$~erg~cm$^{-3}$, whereas other components at different frequencies (X rays, gamma rays, etc.) are much more subdominant by orders of magnitude.
Therefore, the supernova neutrinos, even though we have not seen them yet, make up of one of the brightest radiation components in the Universe.

We can reach the same conclusion by another simple consideration.
For each galaxy, its time-averaged luminosity in supernova neutrinos is $L_{\rm DSNB,gal} = E_b  R_{\rm SN,gal} = 2\times 10^{44}$~erg~s$^{-1}$, where $R_{\rm SN,gal} = 0.02$~yr$^{-1}$ is the supernova rate per galaxy.
On the other hand, the typical luminosity of galaxies (in star light) is $L_{\star,{\rm gal}} = 10^{10}L_{\odot} = 4\times 10^{43}$~erg~s$^{-1}$.
Therefore, if we average over a long time scale (such as Hubble time), then the Milky-Way Galaxy shines brighter with supernova neutrinos than its optical photons emitted by stars by a factor of a few. 
We must, therefore, find this component observationally.

\subsection{Kinetic equation}

The differential intensity of the neutrino beam $I(E,t)$ --- the number of neutrinos received per unit area, unit time, unit energy range, and unit solid angle --- is the most relevant quantity for the DSNB.
It is related to the number density per unit energy range, $n(E)$, through $I(E) = n(E) / (4\pi)$, where we adopt the natural unit of $c = \hbar = 1$.
Because of the homogeneity and isotropy of the Universe, the differential number density of the neutrinos, $n(E)$, is independent of spatial coordinates but is a function of the neutrino energy $E$.
The current and near future detectors will have no directional sensitivity to enable study of the DSNB anisotropies, and thus the differential flux $F(E) = 4\pi I(E) = n(E)$ has been used in the literature.

We want to understand the time evolution of the differential number density of the DSNB, $n(E,t)$, which we define as a number of the neutrinos per {\it comoving} volume per unit energy range.
It is related to the phase space density $f(E,t)$ through,
\begin{equation}
    n(E,t) = \frac{E^2}{2\pi^2}f(E,t)a^3(t),
\end{equation}
where $a(t)$ is the scale factor of the Universe, which converts the physical density to the comoving density. The phase space density $f(E,t)$ depends on $(E,t)$ as the distribution is assumed homogeneous and isotropic, and neutrinos are relativistic, $E = |\vec p|$.

The evolution of $f(E,t)$ is described by the following kinetic equation:
\begin{equation}
    \hat L[f(E,t)] = \mathcal C[f(E,t)],
\end{equation}
where $\hat L$ is the Liouville operator and $\mathcal C$ is the collisional term. 

The Liouville operator $\hat L$ in generic spacetime is
\begin{equation}
    \hat L = p^\alpha \frac{\partial}{\partial x^\alpha}-\Gamma^\alpha_{\beta\gamma}p^\beta p^\gamma\frac{\partial}{\partial p^\alpha},
\end{equation}
where $\Gamma^\alpha_{\beta\gamma}$ is the Christoffel symbol.\cite{Kolb:1990vq} By applying this to the Friedmann-Robertson-Walker metric, we have
\begin{equation}
    \hat L[f(E,t)] = E\frac{\partial f}{\partial t}-H(t)E^2\frac{\partial f}{\partial E},
\end{equation}
where $H(t) = \dot a(t)/a(t)$ is the Hubble function.
Rewriting this in terms of $n(E,t)$, then, the Liouville operator acting on $n(E,t)$ is
\begin{equation}
    \hat L[n(E,t)] = E\left[\frac{\partial}{\partial t} -H(t) E\frac{\partial}{\partial E}-H(t)\right]n(E,t).
\end{equation}
The ``collisional'' term on the right-hand side is then given by
\begin{equation}
    \mathcal C[n(E,t)] = R_{\rm SN}(t) E\frac{dN(E)}{dE},
\end{equation}
where $R_{\rm SN}(t)$ is the supernova rate per comoving volume, and $dN/dE$ is the neutrino spectrum per each supernova.
To summarize, the neutrino kinetic equation is
\begin{equation}
    \left[\frac{\partial}{\partial t} -H(t) E\frac{\partial}{\partial E}-H(t)\right]n(E,t) =  R_{\rm SN}(t) \frac{dN(E)}{dE},
    \label{eq:neutrino kinetic equation}
\end{equation}

\subsection{Solution}

We largely follow Ref.~\cite{Fogli:2004gy} for solving the kinetic equation.
First, instead of $(E,t)$ variables, we adopt $(\varepsilon,z)$, where $z$ is the cosmological redshift and $\varepsilon = E/(1+z)$.
Then, the derivative operators $\partial_t$ and $\partial_E$ can be written as $\partial_t = -H(z)[(1+z)\partial_z - \varepsilon \partial_\varepsilon]$ and $\partial_E = (1+z)^{-1}\partial_\varepsilon$.
With these new variables, Eq.~(\ref{eq:neutrino kinetic equation}) can be rewritten as
\begin{equation}
-H(z)\frac{\partial}{\partial z}\left[(1+z)n(\varepsilon, z)\right] = R_{\rm SN}(z)\left.\frac{dN}{dE}\right|_{E=(1+z)\varepsilon},
\end{equation}
which can be integrated to obtain
\begin{equation}
    n(\varepsilon,z) = \frac{1}{1+z}\int_z^\infty \frac{dz'}{H(z')}R_{\rm SN}(z')\left.\frac{dN}{dE}\right|_{E=(1+z')\varepsilon}.
\end{equation}
We are interested in the differential flux of the neutrinos at $z = 0$, and since $F(E) = n(E,z=0)$, we finally obtain
\begin{eqnarray}
    F(E) &=& \frac{c}{H_0}\int_0^\infty \frac{dz}{\sqrt{\Omega_m(1+z)^3+\Omega_\Lambda}}
    \nonumber\\&&{}\times
    R_{\rm SN}(z) \left.\frac{dN}{dE'}\right|_{E' = (1+z)E}.
\end{eqnarray}
Here we note that we recovered $c$ explicitly and used the Friedmann equation in the flat expanding Universe, $H(z) = H_0\sqrt{\Omega_m(1+z)^3+\Omega_\Lambda}$, which is dominated by matter and the cosmological constant $\Lambda$ with their density parameters, $\Omega_m$ and $\Omega_\Lambda$, respectively.
Throughout this review, we adopt $H_0 = 67.4~\mathrm{km~s^{-1}~Mpc^{-1}}$, $\Omega_m = 0.315$, and $\Omega_\Lambda = 0.685$, which are compatible with the latest Planck measurements of the cosmological parameters.\cite{Planck:2018vyg}

\section{Theoretical model ingredients}

\subsection{Cosmic supernova rate density}

\subsubsection{Direct measurements}\label{sec:directmeasurements}

While supernovae have been directly observed for centuries, systematic observations which allow their volumetric rates to be inferred has been driven more recently by the rise of large surveys. Since supernovae are transient phenomena, their search requires multiple scans of the same target(s). There are broadly two survey strategies to search for supernovae: targeted and non-targeted. 

In the targeted strategy, a list of targets (e.g., galaxies) is pre-compiled and surveyed repeatedly over time. Early surveys\cite{Cappellaro:1999qy,Cappellaro:2004ti,Botticella:2007er} often adopted this strategy because the number of discovered supernovae can be maximized by targeting large star-forming galaxies. Such surveys therefore often report supernova rates per unit galaxy luminosity in a particular band, e.g., SNuB = 1 SN $(100 \, {\rm yr})^{-1} (10^{10} L_\odot^{\rm B})^{-1}$ where $L_\odot^{\rm B}$ is the solar B-band luminosity. To convert this to the \emph{volumetric} rate needed for the DSNB, one must account for the luminosity density of the Universe at the redshift in question which need to be independently measured, e.g., $j_B(z) = (1.03 + 1.76 z) \times 10^8 L_\odot^B \,{\rm Mpc}^{-3}$\cite{Botticella:2007er}. One concern is that targeted surveys would miss bright supernovae in smaller galaxies not included in the target list, even if they are within the survey's flux sensitivity and volume range. This must be corrected for. In the more recent survey by the Lick Observatory Supernova Search (LOSS)\cite{Leaman:2010kb,Li:2010kc,Li:2010kd}, over 10,000 galaxies were monitored over the course of 11 years, leading to the discovery of the ``rate-size relation'' where the supernova frequency is not linearly proportional to the size of the host galaxy, but as a power law with an index (of 0.4–0.6), whose exact number depends on the supernova type and galaxy Hubble type. Thus, correcting for missing supernovae is far from a simple scaling exercise. 

In the non-targeted strategy, patches of the sky are pre-selected and surveyed repeatedly over time. This locates all supernovae only limited by flux, i.e., regardless of the supernova host galaxy. This means the volumetric rate is more readily estimated. On the other hand, a large field of view must be surveyed to acquire a large sample of supernovae, especially for nearby distances. Nevertheless, most recent surveys have adopted this method. The Rubin Observatory Legacy Survey of Space and Time (LSST), for example, is expected to survey $\sim20,000\;\mathrm{deg}^2$, slightly less than half of the sky, with a planned revisit time of $\sim 3$ days on average per 10,000 deg$^2$ with 2 visits per night.\cite{LSST:2008ijt} This will result in orders of magnitude increase in the number of supernovae discovered \cite{Lien:2009db,LSST:2008ijt}.

An important systematic consideration affecting both survey strategies is that of dust extinction. Due to flux sensitivity limits, supernova surveys will miss the faintest supernovae\cite{Mattila:2012zr}. Over a population, the supernova luminosity has some intrinsic distribution, so it is natural some will fall below the survey sensitivity; but intrinsically luminous supernovae can also appear significantly fainter due to dust extinction. Typically, the supernova luminosity function is constructed using a volume-limited sample (for example, either obtained independently or using a sub-set of nearby supernovae within the survey) and used to infer the missing faint end in larger flux-limited samples. However, dust extinction is redshift-dependent and its modeling for supernovae correction has been updated over the last decades.\cite{Mannucci:2007ex,Mattila:2012zr} As explained later in Section \ref{sec:BHrate}, it has been pointed out that the direct supernova measurements are systematically lower than the birth rates of massive stars\cite{Horiuchi:2011zz}. This discrepancy between the birth rates and supernova rates can be remedied by such improved dust modeling. 

Furthermore, the DSNB is sensitive to the total core collapse rate, regardless of whether the collapse yields a luminous supernova or not. In this context, direct measurements of supernovae is likely to fall short, since the collapse to black holes are expected to yield systematically dimmer, longer, and redder transients\cite{Lovegrove:2013ssa,Lovegrove:2017trg} which are not efficiently detected by supernova surveys\cite{Gerke:2014ooa,Neustadt:2021jjt,Byrne:2022oik}.

\subsubsection{Measurements of birth rates}

An alternative method to measure the cosmic supernova rate is indirectly from the cosmic birth rate of stars: the {\it mass} of the total stars formed per unit time per unit comoving volume at a given redshift $z$. This strategy works since the lifetimes of massive stars which undergo supernova explosions (heavier than approximately $\sim 8 M_\odot$ in mass) is less than $O(100)$ Myr which is short on cosmological timescales. In other words, all massive stars born in a particular redshift range will expend their nuclear fuel and collapse within the same redshift bin, without a cosmologically significant delay (the same cannot be said about thermonuclear supernovae\cite{Horiuchi:2010kq}, which can be delayed by billions of years since the birth of their progenitor stars). 

The core-collapse rate density is then inferred as follows assuming the stellar initial mass function (IMF) $\phi(M)$:
\begin{equation}
    R_{\rm CC}(z) = \frac{\int_{8M_\odot}^{125M_\odot} dM\phi(M)}{\int_{0.1M_\odot}^{125M_\odot} dM M\phi(M)}R_{\rm SF}(z),
\end{equation}
where $R_{\rm SF}$ is the star formation rate density, and the Salpeter mass function ($\phi(M) \propto M^{-2.35}$) has been adopted often in the literature\cite{Salpeter:1955it}. Here, the stellar mass distribution spans $0.1 M_\odot$ to $125 M_\odot$, with the supernova progenitor ranging from $8 M_\odot$ to $125 M_\odot$. In reality, the Salpeter IMF is known to be inaccurate especially at the low-mass end, but it is still useful as a common way to compare. Various modifications to the Salpeter IMF exist, among the more frequently used including those by Chabrier\cite{Chabrier:2003ki} and Kroupa.\cite{Kroupa:2000iv}

\begin{figure}[h]
    \centering
    \includegraphics[width=\linewidth]{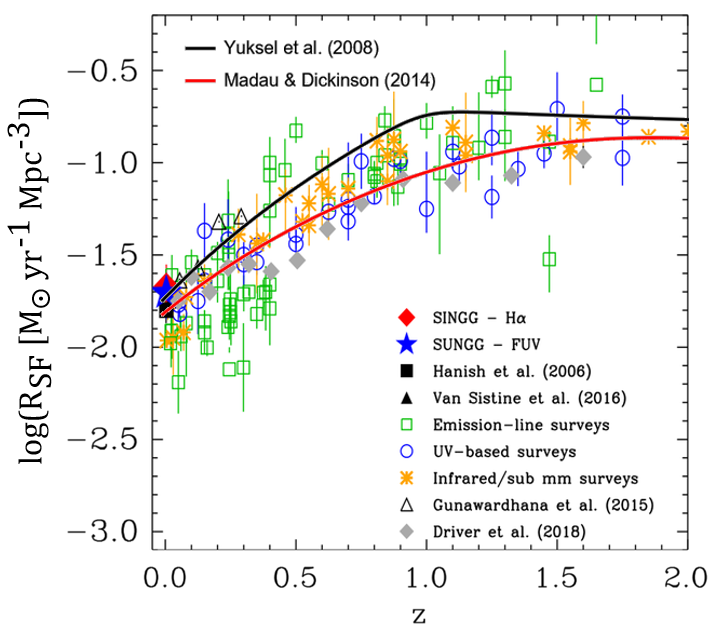}
    \caption{Star formation rate density (SFRD) as a function of redshift. SFRD points collected from the right panel of Figure 6 of Audcent-Ross et al. 2018\cite{Audcent_Ross_2018}, entitled ``Near-identical star formation rate densities from H$\alpha$ and FUV at redshift zero'', published in MNRAS volume 480 no. 1. Fits to other data sets performed in Madau and Dickinson 2014\cite{Madau:2014bja} and Yuksel et al. 2008\cite{Yuksel:2008cu} with Horiuchi et al. 2008\cite{Horiuchi:2008jz} parameters are plotted for comparison. All assume the Salpeter IMF.} 
    \label{fig:starformationrates}
\end{figure}

The cosmic birth rate of stars is inferred from the number of massive stars in a galaxy at a given time. From this, the mean star formation rate of stars during the time window equivalent to the lifetimes of massive stars can be estimated. It is, however, not feasible to observe individual massive stars in distant galaxies. So instead, aggregate galactic observables are used to estimate the number of massive stars: so-called ``star-formation indicators.''\cite{Kennicutt:1998zb} For example, massive stars have high surface temperature and thus typically dominate a galaxy's ultraviolet (UV) emission (barring emission from non-stellar origins, e.g., active galactic nuclei, AGN). Thus, observations of a galaxy's UV emission can be combined with knowledge of massive stars' UV luminosities to estimate the recent star formation rate. The conversion to star formation rate is known as the ``calibration factor'' and its modeling typically relies on stellar population modeling. In practice, when the star-formation rate estimated using, e.g., the UV indicator, the observed UV luminosity is appropriately dust-corrected and AGN corrected (when necessary), then multiplied by the UV calibration factor. A recent collection of star formation rate density measurements is shown in Fig.~\ref{fig:starformationrates} (including UV and other measurements), alongside two commonly used functional fits to the data from Yuksel et al. 2008\cite{Yuksel:2008cu} and Madau and Dickinson 2014\cite{Madau:2014bja}. In the redshift range of a few relevant for the DSNB, the spread in measurements is a factor of 2--3. 

Calibration factors have been computed for the various star-formation indicators but requires a great deal of complex physics, including challenging problems such as the evolution of massive stars and its dependence on stellar parameters, such as mass, metallicity, and rotation; modeling of stellar atmospheres; the binary fraction and the resulting interactions; the shape of the stellar IMF; whether star formation is continuous or variable; and so on. The status was summarized by Kennicutt in 1998,\cite{Kennicutt:1998zb} which continues to be used for ease of comparison with other estimates, but have also been updated for various specific indicators and circumstances. 

One of the significant uncertainties arises from the IMF. Physically, this is because an extrapolation in mass is needed, which in turn depends on the IMF. More specifically, although the measurement of star formation is based on the population of massive stars, the \textit{total} star-formation rate (SFR) is dominated by low-mass stars. Uncertainties in the shape of the IMF introduces a factor $\sim 2$ or more effect on the inferred SFR. Fortunately, this uncertainty does not propagate directly to the DSNB, because the DSNB is not powered by low-mass stars. Instead, the DSNB is powered by the same massive stars used to measure the star formation, thus the effect of the IMF uncertainty on the supernova rate is at the level of $<10$ percent.\cite{Horiuchi:2008jz} This is also corroborated by a study of the implications of a non-Universal IMF, where even extreme IMF variations led to small variations in the supernova rate, at least in the low redshifts of relevance for the DSNB.\cite{Ziegler:2022ivq}

\subsubsection{Core collapse to black holes}\label{sec:BHrate}

As mentioned earlier, the core collapse of a massive star need not necessarily cause a supernova explosion. In fact, the observation of stellar-mass black holes---from microquasars to gravitational waves---motivate that a non-negligible fraction of massive stars should collapse to black holes, creating potentially weak or negligible supernova explosions. However, they are still powerful neutrino sources. 

So far, the in-situ formation of a black hole has not been directly observed. Theoretical investigations suggest the direct collapse of the core of a large red supergiant to a black hole could yield a weak, red, and long-lived transient due to the unbinding of the hydrogen envelope\cite{Lovegrove:2013ssa,Lovegrove:2017trg}. However, transient surveys are not tuned to capture such long and red explosions. Alternatively, if an explosion is hard to observe, the disappearance of a massive stars might be easier. Such a survey has in fact been in operation for about a decade. This is the ``survey about nothing,'' searching for the disappearance of massive stars without a supernova explosion\cite{Kochanek:2008mp}. These stars are then strong candidates for stars which undergo core collapse into black holes. The general strategy is to monitor a sufficient number of massive stars such that on average one of them will core collapse in any given year. In 11 years running, the survey has 9 massive stars coincident in time and location with supernovae, and 2 massive stars disappearing without such coincident supernovae\cite{Gerke:2014ooa,Adams:2016ffj,Adams:2016hit,Basinger:2020iir,Neustadt:2021jjt}. Naively, this implies a $23.6^{+23.3}_{-15.7}$\% fraction of massive stars undergoing collapse to a black hole; note that if only one of the disappearing candidate is included, the fraction is $16.2^{+23.2}_{-12.5}$\%\cite{Neustadt:2021jjt}. Even for the more conservative latter estimate, it is still a substantial enough fraction to be of interest for the DSNB. While the uncertainty is large, this is a non-negligible fraction for the DSNB. One of the benefits of using the star formation rate is that it does not bias against collapse to black holes. 

In principle, precisely knowing the cosmic star-formation rate and the cosmic supernova rate independently can yield the fraction of core collapse to black holes. Early measurements suggested a systematic discrepancy between them, at the factor of $\sim 2$ level,\cite{Horiuchi:2011zz} that naively could be interpreted as some half of massive stars are collapsing to black holes without luminous supernovae. However, supernovae could be faint for other reason, chief among them dust extinction (see section \ref{sec:directmeasurements}). Indeed, the authors\cite{Horiuchi:2011zz} found that at nearby distances $< 20$ Mpc, where even dust extinguished supernovae could be more easily detected, the fraction of faint supernovae were much higher than at larger distances, suggesting dust to be an important effect. Since then, updated dust models have been applied to supernova rate measurements, which have bridged the gap between the supernova rate and star-formation rates\cite{Graur:2014bua}. Given the existing uncertainties, fractions of $\sim 10$--$40$\% seem to be allowed, consistent with searches of disappearing stars. 

Theoretically predicting the fraction of stars which collapse to black holes is very uncertain at present, due to the unknown nature of the explosion mechanism and the need to survey a vast landscape of massive star properties. Attempts have been made assuming the delayed-neutrino heating explosion mechanism and using simpler core-collapse simulation treatments to survey large numbers of progenitors. For example, a recent analysis by the Garching group yields fractions ranging from 17.8\% up to 41.7\%\cite{Kresse:2020nto}. As a result, most studies therefore parameterize the fraction of core collapse into black holes. 

\subsection{Neutrino spectrum from core-collapse supernova explosions}

Broadly speaking, in predicting the DSNB we need to account for: (i) the emission of neutrinos from the vast range of properties of massive stars in nature, (ii) the time-integrated neutrino emission, often going beyond the time ranges studied by numerical simulations, (iii) the emission of neutrinos from both core collapse to neutron stars and black holes, and (iv) a consideration for neutrino oscillations in the progenitor. 

\subsubsection{Progenitors}

It is generally accepted that stars heavier than $\sim 8 M_\odot$ end their life triggered by a core collapse, leaving a neutron star or a black hole behind. How the core collapse depends on properties of progenitors such as their mass and rotation has been investigated extensively; however it is still far from complete. For example, the conditions for black-hole formation after the core collapse was once considered to depend mostly on the mass and metallicity of its progenitor star, with massive and low metals being the important criteria\cite{Woosley:2002zz}. This implied for galaxies dominated by solar metal stars, the fraction of massive stars collapsing to black holes was only a few percent.\cite{OConnor:2010moj} It turned out, however, to be not so simple. Shock revival is highly sensitive to the energy transport by neutrinos, which in turn depends on both the microphysics governing the emission from the hot dense core and the three-dimensional turbulence structure determining the neutrino capture (and hence heating) rate. In other words, whether an explosion occurs is sensitive to the initial conditions which appear to be not simply monotonic with e.g., the stellar mass.\cite{Ugliano:2012fvp,Pejcha:2014wda,Nakamura:2014caa,Sukhbold:2015wba,Kresse:2020nto}  As a result, black-hole formation and hence a failed supernova can happen even for low-mass progenitors as light as $\sim 15M_\odot$, while very heavy stars can still cause successful explosions.

On the other hand, the neutrino emission can be more accurately characterized. For example, the stellar compactness parameter,\cite{OConnor:2010moj}
\begin{equation}
    \xi_M = \frac{M/M_\odot}{R(M_{\rm bary}=M)/\mathrm{1000\,km}},
\end{equation}
has been discussed in the literature. Here, $R(M_{\rm bary}=M)$ is a radius enclosing a baryonic mass $M$. The compactness at $M = 2.5M_\odot$, $\xi_{2.5}$ is a suitable criteria for the black-hole formation since it is close to the maximal mass of a neutron star\cite{OConnor:2010moj}, although other masses can provide clearer distinguishing power\cite{Nakamura:2014caa}. Reference~\cite{Horiuchi:2017qja} compiled core-collapse simulations of 8--$100M_\odot$ progenitor masses, and found extensive dependence of neutrino emission spectrum on $\xi_{2.5}$. Approximately, a higher compactness star has a large massive core, hence a longer lasting intense mass accretion onto the collapsed core. This accretion powers strong neutrino emissions, and thus both the total energy liberated in neutrinos and the neutrino mean energy grows with compactness. However, at some stage, the mass accretion would become too intense and the shock revival cannot be achieved by the neutrinos. This may suggest a critical compactness necessary for collapse to black holes. In reality, this simplistic picture is complicated by many factors, which smears the predictability noticeably. For example, the progenitor density shift at the silicon shell appears to play an important role in shock revival\cite{Tsang:2022imn}, which is not sufficiently captured in a simple compactness picture. 

Another important consideration for supernova progenitors is binary effects. Observations of stars in the Milky Way and nearby stars reveal that the majority (>50\%) of massive stars have experienced binary effects.\cite{Sana:2012px} Binary interactions, especially mass transfer and stellar mergers, can significantly change the masses of core-collapse progenitors and strongly influence the DSNB. For example, stars which are born below the core collapse threshold ($\sim 8 M_\odot$) may eventually exceed the threshold by, e.g., merging with another star. By adopting population synthesis of binary-star systems, binary effects were found to enhance the DSNB flux by 15--20\% in favor of future detection.\cite{Horiuchi:2020jnc}

\subsubsection{Neutrino emission}

Neutrinos are emitted from the surface of the protoneutron star.
Therefore, neutrinos---along with gravitational waves---is one of the only ways to probe the condition of the collapsing core. To date, the only observational dataset is the approximately two dozen neutrinos from SN1987A\cite{Kamiokande-II:1987idp,Bionta:1987qt}; thus, for predicting the DSNB we must in general rely on results of numerical simulations of stellar core collapse.
They are, however, one of the most challenging problems in computational astrophysics\cite{Kotake:2005zn,Mirizzi:2015eza,Janka:2017vlw,Burrows:2020qrp}, as they involve all the four fundamental forces of nature (electromagnetic, gravitational, strong, and weak forces) playing key roles and one must address microphysics of neutrino propagation, with gravito-magnetohydrodynamics part of the code.
Especially, for obtaining the most realistic outcomes, we should perform the simulations in the three-dimensional setup, including both the coordinate space and neutrino momentum space, which makes it computationally expensive. Nevertheless, researchers from multiple groups have tackled this problem and have reached reliable results.
For example, they now successfully lead to supernova explosions especially for low-mass progenitor models.
Recent three-dimensional simulations of the $\sim 20M_\odot$ progenitor with elaborate neutrino transfer scheme successfully obtained the explosion energy of $\sim 10^{51}$\,erg that matches supernova observations well through neutrino-heating mechanism.\cite{Bollig:2020phc}

Despite this incredible progress in the core-collapse simulations, most simulations are still limited to within the initial one second or so. This is mainly due to the fact that successful shock revival must occur during the initial second while the neutrino emission is most intense. However, the time-integrated spectrum is relevant for the DSNB flux estimate, and indeed, a substantial fraction of the total neutrino emission attributes to the later phase of the protoneutron star cooling after one second (typically $>50\%$). We discuss this deficit in detail in Section \ref{sec:latetime}.

\begin{figure}[h]
    \centering
    \includegraphics[width=\linewidth]{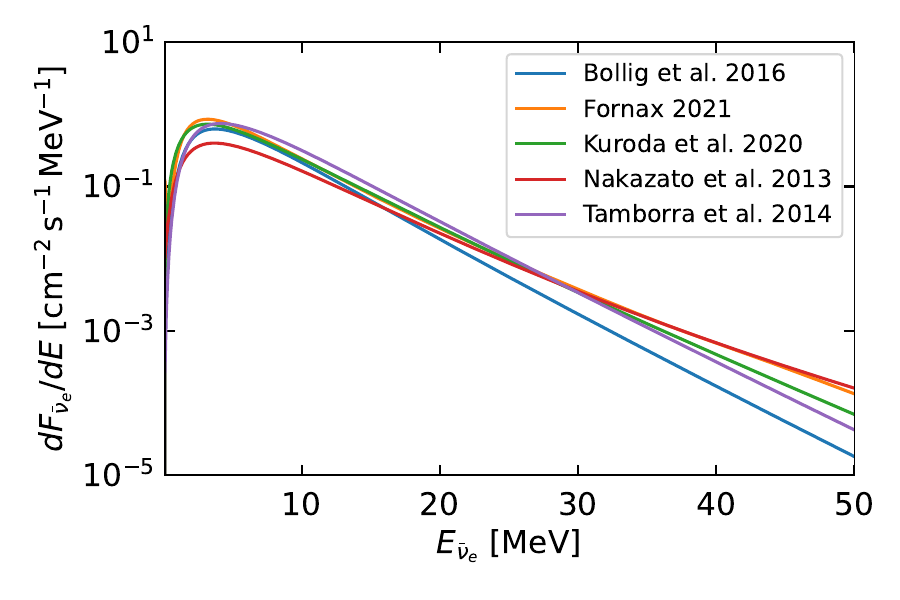}
    \caption{DSNB spectra of different models for the hydrodynamic phase of protoneutron star evolution; each model shown here has been integrated up to the first $300\,{\rm ms}$ of each simulation for $20\,M_{\odot}$ progenitors (except for Bollig et al. 2016\cite{Mirizzi:2015eza} which is a $27\,M_{\odot}$ progenitor and the Nakazato et al. 2013\cite{Nakazato:2012qf} model has metallicity $Z=0.02$). We compare these against the Fornax 2021\cite{Burrows:2020qrp,Nagakura:2021lma}, Kuroda et al. 2020\cite{Kuroda:2020pta}, and Tamborra et al. 2014\cite{Tamborra:2014hga} models.}
    \label{fig:hydro_comparison}
\end{figure}

Multiple independent groups compared one-dimensional results from each of their simulations, and confirmed that, even with different approximations, etc., the results are in good agreement.
For the neutrino emission, the agreement level is within a few tens of percent.\cite{OConnor:2018sti} This is confirmed as well in Fig.~\ref{fig:hydro_comparison}, where we compare the time-integrated spectra from five simulation models of a $20\,M_{\odot}$ progenitor (with the exception of Bollig et al.\cite{Mirizzi:2015eza}, which is a $27\,M_{\odot}$ progenitor) up to the first $300\,{\rm ms}$ post core collapse.\footnote{All DSNB flux spectra figures are created using our publicly available PyDSNB code, found at \url{https://github.com/shinichiroando/PyDSNB/tree/main}. This code makes use of and is consistent with the most recent version of SNEWPY (version 1.3).\cite{Baxter2021,SNEWS:2021ewj}}

Since the neutrino spectrum from each core-collapse event depends on characteristics of progenitor stars such as their masses, for the estimate of the DSNB flux, one must take the average of $dN/dE$ over mass with the weight of the stellar IMF:
\begin{equation}
    \frac{dN}{dE} = \frac{\sum_i(dN/dE)_{i} \int_{M_i^l}^{M_i^u} dM\phi(M)}{\int dM\phi(M)}.
\end{equation}
Here one can use a model spectrum $(dN/dE)_i$ in a mass bin $i$ between the lower and upper mass bounds, $M_i^l$ and $M_i^u$, respectively.

\subsubsection{Neutrino oscillations}

It is strongly established that the neutrinos have masses and different flavors mix during propagation. 
These phenomena called neutrinos oscillations are well understood in vacuum and matter-dominated environments, particularly because all the mixing angles and mass squared differences have been measured precisely. 
There is still uncertainty related to the mass hierarchy that can be either normal (NH) or inverted (IH).
The MSW mechanism of the matter-induced neutrino oscillations mixes the electron flavors ($\nu_e$ and $\bar\nu_e$) with heavy-lepton flavors ($\nu_x$ collectively representing $\mu$ and $\tau$ flavor neutrinos and anti-neutrinos) as follows:
\begin{equation}
F_{\nu_e} =\left\{
\begin{array}{lll}
 F_{\nu_x}^0 & \mbox{(NH)}, \\
 \sin^2\theta_{12} F_{\nu_e}^0 + \cos^2\theta_{12} F_{\nu_x}^0 & \mbox{(IH)}, 
\end{array}\right.
\end{equation}
for the neutrino sector, and
\begin{equation}
F_{\bar\nu_e} = \left\{
\begin{array}{lll}
\cos^2\theta_{12} F_{\bar\nu_e}^0 + \sin^2\theta_{12} F_{\nu_x}^0 & \mbox{(NH)}, \\
F_{\nu_x}^0 & \mbox{(IH)},
\end{array}\right.
\end{equation}
for the anti-neutrino sector.
Here, $\theta_{12}$ is an mixing angle that is relevant for solar and reactor neutrinos ($\sin^2\theta_{12} \approx 0.3$~\cite{ParticleDataGroup:2022pth}). For simplicity, we assume $\theta_{13} = 0$. Using a more accurate value ($\sin^2\theta_{13} \approx 0.02$~\cite{ParticleDataGroup:2022pth}) will have a negligible impact on the obtained results. The superscript 0 represent the flux at production (i.e., before propagation through the stellar envelope and interstellar space).

\begin{figure}
    \centering
    \includegraphics[width=\linewidth]{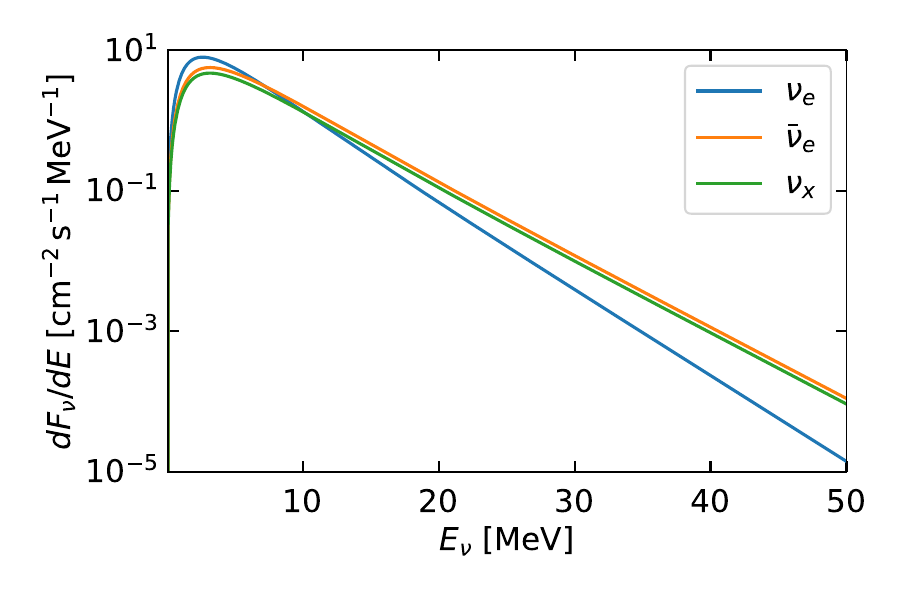}
    \caption{DSNB spectra for the Fermi-Dirac model with parameters derived from the model of Ekanger et al.\cite{Ekanger:2022neg} which have $E_{\nu,e} = 5.8\times10^{52}\,{\rm erg}$, $E_{\bar\nu,e} = 6\times10^{52}\,{\rm erg}$, and $E_{\nu,x} = 5\times10^{52}\,{\rm erg}$, and temperatures $T_{\nu,e} = 3.4\,{\rm MeV}$ and $T_{\bar\nu_e} = T_{\nu,x} = 4.1\,{\rm MeV}$. Because $\bar\nu_e$ and $\nu_x$ have very similar temperatures, the spectrum after oscillation due to the MSW effect looks very similar for normal and inverted hierarchies.} 
    \label{fig:flavor_contributions}
\end{figure}

Around the neutrinosphere where the density of the neutrinos is large, collective oscillations triggered by the neutrino-neutrino interactions can be important. 
This is likely to be important in a time-dependent manner for example in the next Milky Way core-collapse supernova, but could be smeared out in the DSNB spectrum where the signature is integrated over time.\cite{Lunardini:2012ne, Moller:2018kpn} However, the jury is still out given the dearth of predictable collective oscillation patterns. We thus neglect the effect of collective neutrino oscillation on the DSNB flux, and focus only on the MSW effect. In Fig.~\ref{fig:flavor_contributions}, we show the DSNB flux spectra for each flavor with a Fermi-Dirac distribution at temperatures of 3.4, 4.1, and $4.1\,{\rm MeV}$ for $\nu_e$, $\bar\nu_e$, and $\nu_x$, respectively, based on the analysis done in Ekanger et al.\cite{Ekanger:2022neg}. Because the $\bar\nu_e$ and $\nu_x$ spectra are so similar, the spectrum after MSW oscillation looks very similar for normal and inverted hierarchies.

\subsubsection{Late-time evolution}\label{sec:latetime}

\begin{figure}
    \centering
    \includegraphics[width=\linewidth]{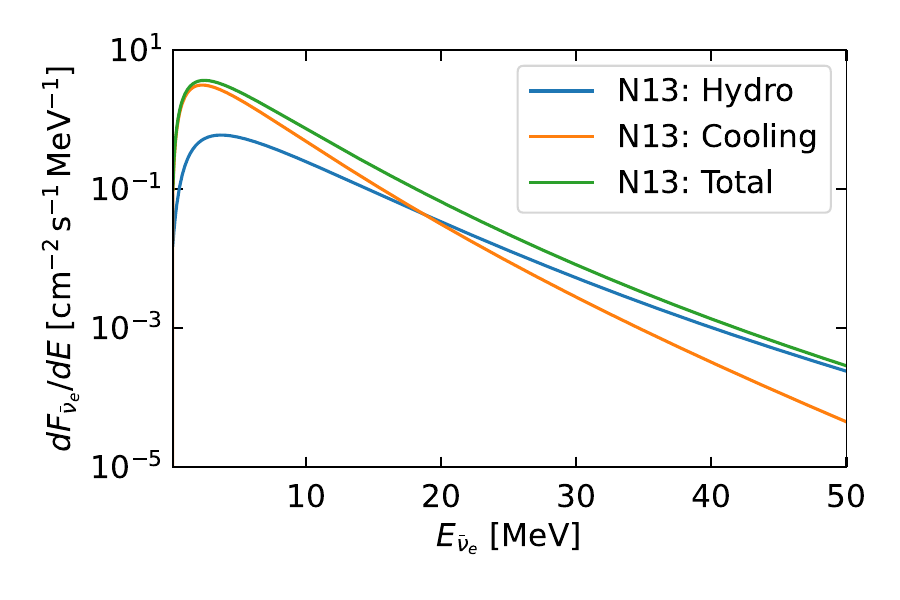}
    \caption{Comparison of the DNSB spectrum from the early hydrodynamic, late cooling, and total phases of the Nakazato et al.\cite{Nakazato:2012qf} (or N13) model. This assumes a revival time of $300\,{\rm ms}$, metallicity of $Z=0.02$, and progenitor mass of $20\,M_{\odot}$.} 
    \label{fig:n13_phases}
\end{figure}

\begin{figure}[h]
    \centering
    \includegraphics[width=\linewidth]{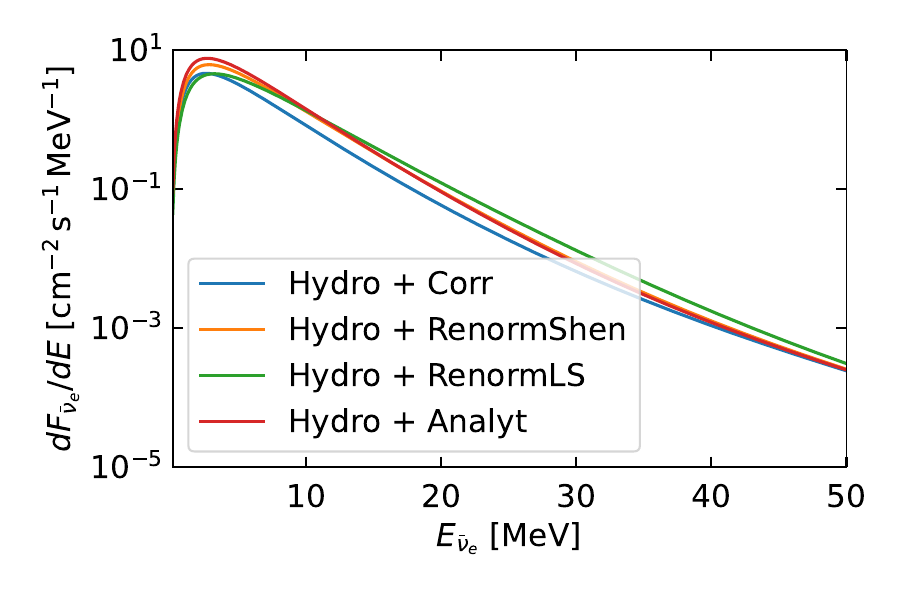}
    \caption{Comparison of the DSNB spectrum of the sum of early hydrodynamic phase plus four different late phase strategies. The data from the hydrodynamic phase is from N13 in this figure, using the same revival time, metallicity, and progenitor mass as Fig.~\ref{fig:n13_phases}.}
    \label{fig:late_phase_comparison}
\end{figure}

Approximately half of the total neutrinos emitted by a stellar core collapse is during the so-called cooling phase, when the protoneutron star cools and shrinks by the emission of mostly neutrinos. Simulations dedicated to the cooling phase have been performed, e.g., exploring the impacts of the protoneutron star's interior turbulence, the equation of state of hot-dense matter, and other nuclear physics. However, due to the considerably different scales from the core-collapse phase, it is not easy to simulate the cooling phase in the same code as the core-collapse phase. Instead, groups have tied together such simulations in various ways. For example, Nakazato et al.\cite{Nakazato:2012qf} performed an analytic ``match'' using a user-defined explosion time. In Fig.~\ref{fig:n13_phases}, we show the DSNB flux spectra for the $20\,M_{\odot}$, $Z=0.02$, $300\,{\rm ms}$ revival time model from Nakazato et al.\cite{Nakazato:2012qf} broken into the early hydrodynamical phase, the late cooling phase, and the total spectrum. The later cooling phase is characterized by lower energy neutrinos and contributes significantly to the overall spectrum. Recently, Ekanger et al.\cite{Ekanger:2022neg} implemented different schemes to model the cooling phase neutrino emissions on the DSNB, and found a factor of $\sim 2$ difference in DSNB rates between them. Fig.~\ref{fig:late_phase_comparison} shows the differences in flux spectra when combining the early hydrodynamical phase from Nakazato et al.\cite{Nakazato:2012qf} and estimating the late phase with four different schemes. While the estimated neutrino luminosities were different between the schemes implemented, the differences in neutrino energies had a larger impact on the predicted DSNB rates. 


\subsubsection{Core collapse to black holes}
\label{sec:Core collapse to black holes}

The neutrino emission from a core collapse to a black hole is expected to be systematically different to that from a core collapse to a neutron star. To understand why, consider that the ultimate energy source of neutrino is gravitational binding energy released as the stellar core collapses to a more compact object. A collapse to a black hole, which has larger mass/radius than a neutron star, therefore releases more energy. More energy density  equates to higher temperatures and luminosities. In reality, the situation is complicated by the time evolution of the core and the fact that once a black hole is formed, the neutrinos from within the event horizon cannot escape and therefore do not contribute to the total neutrino emission. As a result, simulations generally find that compared to collapse to neutron stars, collapse to black holes show (i) higher neutrino energies, especially the heavy-lepton flavor neutrinos whose neutrinospheres are smallest and subject to interior temperatures the most, and (ii) the time-integrated energy released as neutrinos decreases, due to neutrinos not being able to escape. 
A quicker collapse to a black hole exacerbates these features~\cite{Horiuchi:2017qja}. 

How long a certain core will take to collapse to a black hole depends on both the progenitor and on the equation of state of hot-dense matter. For a given equation of state, it is well described by the progenitor's core compactness~\cite{OConnor:2010moj,Horiuchi:2017qja}. While the progenitor compactness in large part sets the mass accretion rate and thus the evolution of the mass of the protoneutron star, the equation of state sets the maximum mass of the collapsing central compact object that can be supported. This in turn depends on the stiffness of the equation of state, the amount of trapped leptons, as well as the temperature in the accreting protoneutron star. While various equations of state models have been explored over the years, the addition of more neutron star mass/radii measurements, as well as the discovery of a neutron-star merger, has helped to narrow the range of possibilities~\cite{Raaijmakers:2021uju}.

\begin{figure}[h]
    \centering
    \includegraphics[width=\linewidth]{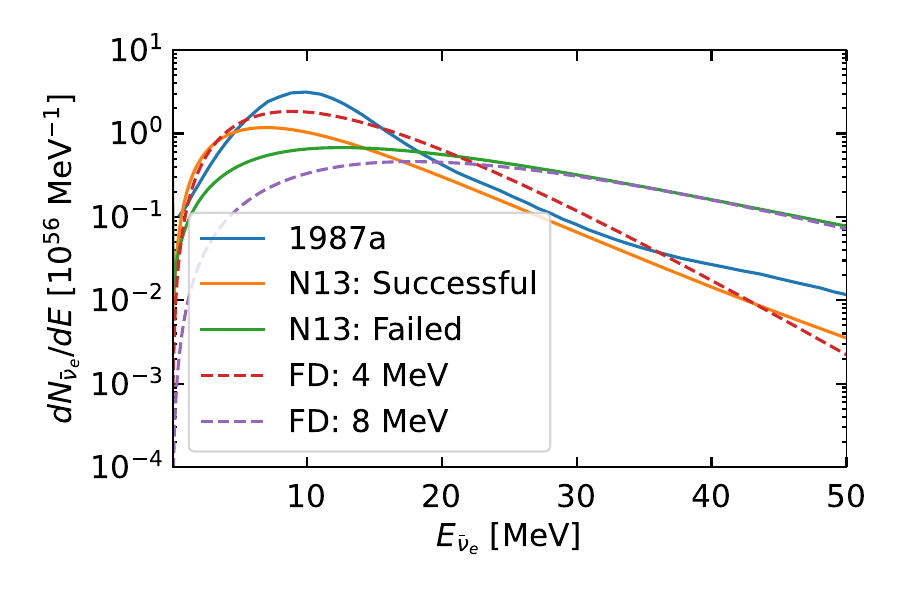}
    \caption{Neutrino number spectrum of successful and failed models of Nakazato et al. (N13)\cite{Nakazato:2012qf} and of SN 1987A.\cite{Yuksel:2007mn} Here, successful N13 model is of $20\,M_{\odot}$ progenitor, revival time is $300\,{\rm ms}$, metallicity is 0.02, and the equation of state is Shen\cite{Shen:1998gq}. Failed model is $30\,M_{\odot}$ progenitor, with 0.004 metallicity, and the equation of state is Shen. Also plotted are Fermi-Dirac distributions for 4 and $8\,{\rm MeV}$ ($3.2\times10^{53}\,{\rm erg}$ total energy liberated) to represent common estimations of successful and failed supernovae, respectively.}
    \label{fig:NSBH1987a_comparison}
\end{figure}

Figure~\ref{fig:NSBH1987a_comparison} shows neutrino spectrum of $\bar\nu_e$ flavor for various models of both successful and failed supernovae. We also compare the models with those reconstructed from the SN~1987A neutrino data as well as conventional Fermi-Dirac distribution (with zero chemical potential) with the temperature of 4~MeV and 8~MeV. If the core collapse leaves a black hole, then the temperature is higher, yielding the neutrino spectrum close to the Fermi-Dirac distribution with 8~MeV temperature.
The core collapse into a neutron star is well approximated as a Fermi-Dirac distribution with 4~MeV temperature.

\section{DSNB spectrum}

\begin{figure}
    \centering
    \includegraphics[width=\linewidth]{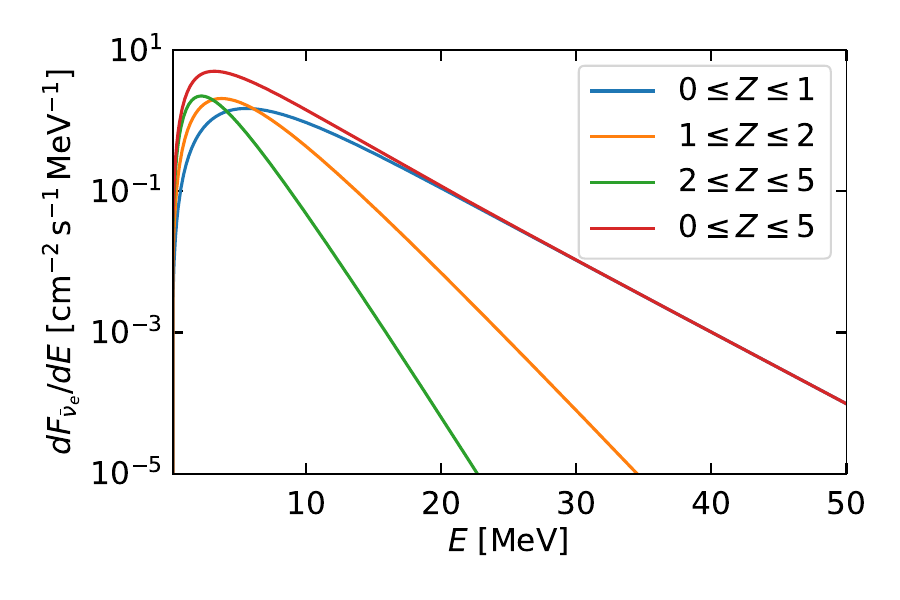}
    \caption{Contributions to the DSNB spectrum from different redshift ranges. The DSNB flux is dominated mostly by supernovae with $z<1$. The Fermi-Dirac distribution with $E_{\rm tot} = 3.2\times 10^{53}\,{\rm erg}$ 
    and $T_{\bar\nu_e} = T_{\nu_x} = 4.1\,{\rm MeV}$ has been adopted.\cite{Ekanger:2022neg}}
    \label{fig:redshift_contributions}
\end{figure}

Because of the redshift of neutrino energies, the DSNB flux is dominated by the contribution from supernovae with $z < 1$ (Fig.~\ref{fig:redshift_contributions}).
At energies above 10~MeV, which is the main target of major detectors such as Super-K, the DSNB spectrum falls exponentially.
Therefore, the DSNB spectrum is often plotted with a linear scale for the horizontal axis representing the neutrino energy and with a linear or log scale for the vertical axis representing the flux.

With the current detection technology, one can adopt neither time nor spatial information, and thus must rely solely on the energy information to probe physics and astrophysics relevant to the DSNB.
Yet, with the high statistics expected with the future generation of neutrino detectors, one can extract various important information on the supernova rate, black-hole formation, and neutrino physics.
We elaborate on each of them in the following.

\subsection{Neutrino spectrum of various flavors}
Even though the dominant detection channel for the DSNB is the inverse beta decay for the $\bar\nu_e$ flavor, it is desirable to detect all the flavors ($\nu_e$, $\bar\nu_e$, and $\nu_x$) to reach a comprehensive understanding of the supernova explosion. We can study $\bar\nu_e$ fluxes with Super-K and Hyper-K, while a good option for the $\nu_e$ flux would be DUNE, which can reach DSNB sensitivity if backgrounds can be controlled.\cite{Tabrizi:2020vmo}
The non-electron flavors $\nu_x$ is the hardest to detect, as one must rely on neutral current interactions.
These are not efficient at MeV energies because either the cross section is suppressed (for neutrino-electron scatterings) or the target is too heavy (for neutrino-nucleon scatterings).
This challenge can be achieved with existing and future dark matter direct-detection detectors such as XENON, LZ, and DARWIN.
Especially with the ultimate dark matter detector, DARWIN, one can reach the sensitivity about an order of magnitude larger than theoretical predictions.\cite{Suliga:2021hek}
It is also possible to adopt the neutral-current interaction off the protons in detectors: $\nu p$ scattering.
It has been studied that if the backgrounds against detecting these scattering events could be well understood and suppressed, one could constrain the $\nu_x$ flavor content using JUNO-like detectors.\cite{Tabrizi:2020vmo} Another suggested opportunity is to use ancient minerals as track detectors, i.e., paleo detectors,\cite{Baum:2023cct} which would record supernova neutrinos over geological timescales. Since the relevant interaction for supernova neutrinos is coherent scattering, all flavor information will be obtained, which can be disentangled by a combined analysis with $\nu_e$ and $\bar\nu_e$ results.\cite{Baum:2022wfc} However, it should be noted that paleo detectors will record more the historical record of Milky Way supernova neutrinos rather than the DSNB.\cite{Baum:2019fqm} This means that while still an average over many supernovae, it is not the true DSNB; the travelled baseline and progenitor samples will be different.

\subsection{Tests of neutrino physics}
Because the DSNB can be regarded as the longest-baseline (at a cosmological distance scale) low-energy (MeV) neutrino experiment, one can probe unique new physics beyond the standard model in the neutrino sector.
Non-radiative neutrino decays induced by the interactions with scalar or pseudo-scalar fields such as Majoron is one such example, where the DSNB can improve upon the sensitivity to neutrino lifetimes by orders of magnitude.\cite{Ando:2003ie,Fogli:2004gy,DeGouvea:2020ang}
Similarly, one can test the pseudo-Dirac nature of neutrinos as it would yield oscillation between the active and sterile species during propagation, and a unique window in the relevant mass-scale parameter $m_k = 10^{-25}$--$10^{-24}$~eV can be tested.\cite{DeGouvea:2020ang}
If the sterile neutrinos at eV mass scales exist and they interact with a new unknown gauge vector boson $\phi$, then such a scenario can be tested using the DSNB.
If $M_\phi = 5$--10~keV and the coupling to the sterile neutrinos $g_s = 10^{-4}$--$10^{-2}$, it would show a characteristic dip in the DSNB energy spectrum that should be captured by the detectors.\cite{Jeong:2018yts}

\subsection{Cosmology and astrophysics}
The DSNB spectrum could also be used to test cosmological models.
By using the DSNB data, one can test the standard $\Lambda$CDM model including the measurement of the Hubble constant\cite{DeGouvea:2020ang} or other exotic models such as a logotropic universe and a bulk viscous matter-dominated universe.\cite{Barranco:2017lug}

\subsection{Core collapse to black holes}

As shown in Sec.~\ref{sec:Core collapse to black holes}, the neutrino spectrum (both the normalization and spectral shape) depends on the final state of collapsed object --- a neutron star or a black hole.
One can use the DSNB spectrum to test the black-hole formation after the core collapse, which is challenging using the optical observations alone.
The flux of models with 100\% BH formation, with a few equations of state~\cite{Ashida:2022nnv}, have already exceeded the 90\% confidence level upper limits.\cite{Super-Kamiokande:2021jaq}
This means that one can already constrain fraction of formation of black holes or high-mass neutron stars.
Future observations with SK-Gd and Hyper-K can constrain these parameters to be smaller than tens of percent levels.\cite{Moller:2018kpn}

\section{Detection prospects}

\subsection{Detection principle and the history of the observation}

There are mainly two types of detectors for DSNB observation. One is the water Cherenkov detector and the other is the liquid scintillator detector.
In a water Cherenkov detector, photo detectors, such as photomultiplier tubes, detect the Cherenkov light produced by charged particles scattered by neutrinos in water.
Cherenkov light is emitted when a charged particle traverses a medium at a velocity greater than the speed of light reduced by the refractive index.
The Cherenkov light is emitted at an angle to the direction of the charged particle due to the refractive index and its velocity, thus, it is detected as a ring image.
If the charged particle is traveling at nearly the speed of light in water, this angle is 42 degrees.
In a liquid scintillator detector, the scintillation light generated by the charged particles as they pass through the material is detected by photo detectors.
Since the light yield is higher than the water Cherenkov detector, it has a lower energy threshold and higher neutron tagging efficiency after neutrino interactions.

The dominant detection reaction with the largest interaction cross section in the DSNB energy region is the inverse beta decay (IBD): $\bar{\nu_e} + p \rightarrow e^+ + n$.
The total cross section is roughly calculated as $\sigma_{\rm tot} \sim 9.52 \times 10^{-44} ({p_e E_e}/{\rm MeV}^2)$~cm$^2$, and the detailed calculations have been published by several authors\cite{STRUMIA200342, PhysRevD.60.053003}.
In this interaction, positron as a prompt signal is detected, in several detectors, followed by gamma rays from neutron capture on proton or nucleus as a delayed signal.
Here, the gamma-ray energy depends on what the neutron is captured by, for example, it is 2.2~MeV in the case of proton capture.
This detection method is called delayed coincidence (DC) and is effective in the reduction of background events compared to measuring only a prompt signal.
In this interaction, the neutrino energy is easily reconstructed from the prompt positron energy using the following equation: $E_{\nu} \sim E_{e^+} + m_n-m_p$, where $m_n$ and $m_p$ are the neutron and proton mass, respectively.

Since the 1980s, DSNB searches have been made with various detectors and, in this section, we briefly introduce each detector and the results.
Unfortunately, all experiments have not observed a significant signal of DSNB and only set a 90\% C.L. upper limit on it.
There are two methods for calculating the limit.
One is the so-called ‘model-independent search’, which is estimated from the number of observed events and the expected background rate without assuming any DSNB or other physics models, which means a flux upper limit of electron antineutrinos.
The other is the so-called ‘model-dependent search’, which is derived by fitting the observed energy spectrum to the spectral shape of the signal from a DSNB model and the background.
Here, we will focus on the results of the model-independent search, i.e., the electron antineutrino flux upper limit. In addition, we will comment on the results of the model-dependent search for some detectors.

The first observation of the DSNB was reported by Kamiokande-II in 1988~\cite{PhysRevLett.61.385}.
It was a 2,140 ton water Cherenkov detector in the Kamioka mine, Japan.
It is famous for the first detection of neutrinos from a supernova explosion~\cite{Kamiokande-II:1987idp}.
As for the DSNB observation, the limit of the electron antineutrinos in the energy region between 19 and 35~MeV was set to 226~cm$^{-2}$~s$^{-1}$ by the exposure of 0.58~kton\,$\cdot$\,yr.

The next observation was reported by LSD in 1992~\cite{AGLIETTA19921}.
It was a liquid scintillation detector, which can tag neutron signals generated via IBD using the DC technique by proton capture.
The upper limit of electron antineutrinos with the exposure of 0.0939~kton\,$\cdot$\,yr was $9.0\times 10^4$ and $8.2\times 10^3$~cm$^{-2}$~s$^{-1}$ for $9 < E_{\nu}/{\rm MeV} < 50$ and $20 < E_{\nu}/{\rm MeV} < 50$, respectively.
They also reported the results of neutrino interactions with carbon nuclei which are sensitive to other flavors of neutrinos although the cross sections are about one order of magnitude lower and the energy thresholds are higher than those of IBD.

\begin{figure}
    \centering
    \includegraphics[width=\linewidth]{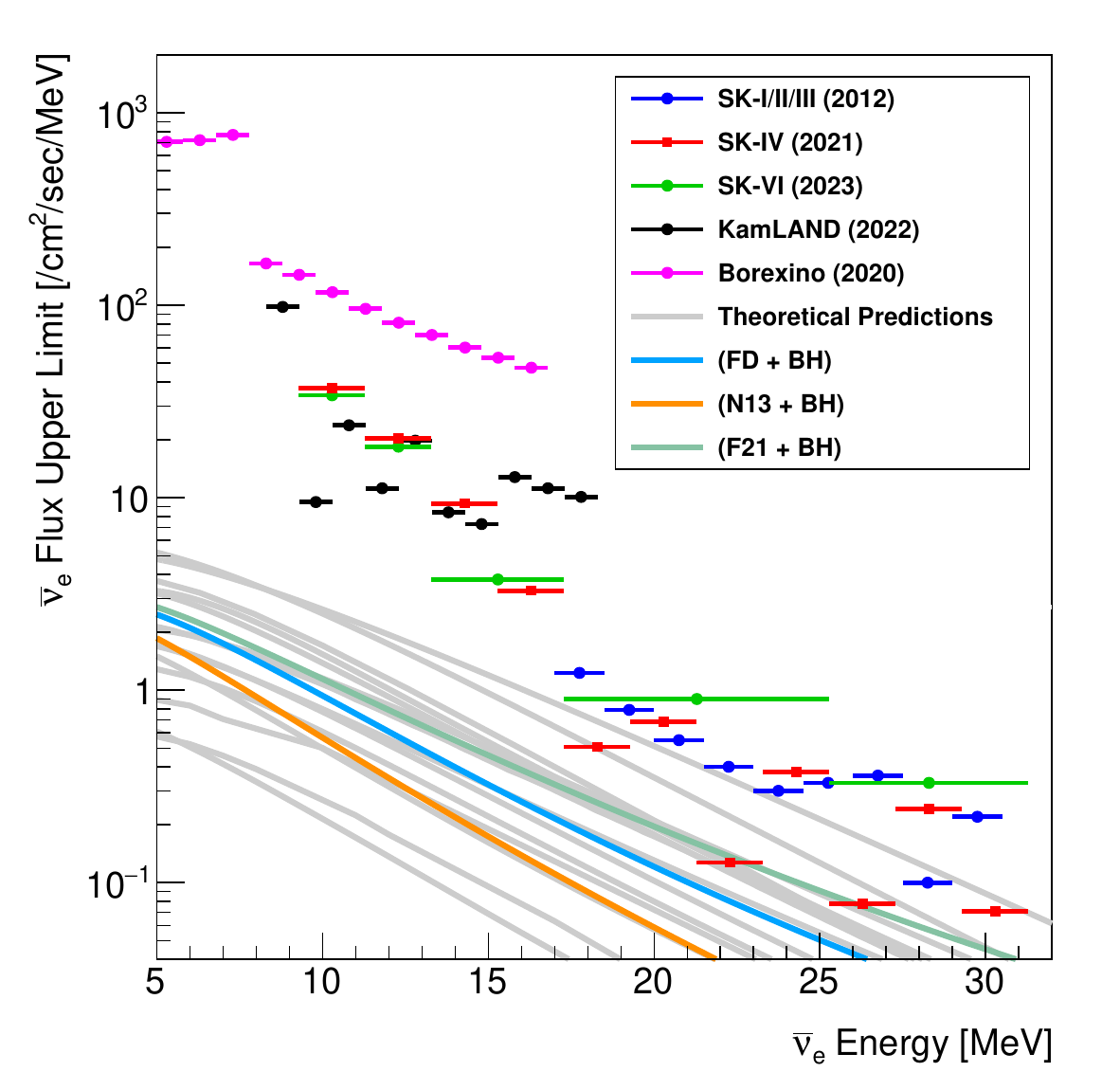}
    \caption{The model-independent upper limits of electron antineutrino flux at 90\% C.L. from various recent experiments, SK-I/II/III~\cite{PhysRevD.85.052007}, SK-IV~\cite{Super-Kamiokande:2021jaq}, SK-VI~\cite{Harada_2023}, KamLAND~\cite{Abe_2022}, and Borexino in the case including atmospheric neutrino background~\cite{AGOSTINI2021102509}, overlaid with various DSNB theoretical predictions. In particular,  three models computed as part of this review are shown in color: a Fermi-Dirac distribution with total energy $3.2 \times 10^{53}$~erg and temperature 4.1~MeV (FD+BH), Nakazato et al.~\cite{Nakazato:2012qf} IMF-weighed spectrum (N13+BH), and Fornax (2021)\cite{Burrows:2020qrp,Nagakura:2021lma} IMF-weighed spectrum (F21+BH). Here, we do not assume a late-phase treatment. A Salpeter IMF is used, the black hole fraction is fixed to be 23.6\%, and the black hole model is from Nakazato et al.~\cite{Nakazato:2012qf} assuming a Shen EOS.}
    \label{fig:flux_limit}
\end{figure}

The situation, that the upper limits are two orders of magnitude higher than the theoretical expectations, has drastically changed since Super-Kamiokande (SK) began in 1996.
It is a 50~kton water Cherenkov detector in the Kamioka mine, Japan.
It is famous for the discovery of neutrino oscillation in atmospheric neutrinos~\cite{PhysRevLett.81.1562}.
The first result of the DSNB observation in SK was reported in 2003~\cite{PhysRevLett.90.061101}.
In this paper, the first phase of SK was used, and the upper limit of electron antineutrinos above 19.3~MeV was 1.2~cm$^{-2}$~s$^{-1}$ by the exposure of 92.2~kton\,$\cdot$\,yr.
The second report appeared in 2012~\cite{PhysRevD.85.052007}, where the combined results of phase I through III in SK with the exposure of 176 kton\,$\cdot$\,yr using only prompt signals were reported.
Figure~\ref{fig:flux_limit} shows the flux upper limit of electron antineutrinos as a function of neutrino energy.
Since the fourth phase of SK (SK-IV) started in 2008, new electronics were installed~\cite{Nishino:2007ccp}.
It enabled the background events reduced by detecting 2.2~MeV gamma-ray from proton capture of a neutron.
SK demonstrated successful detection of the neutron capture signal, although the efficiency was low ($\sim$20\%)~\cite{ZHANG201541}.
The results using the full data set of SK-IV were reported in 2021~\cite{Super-Kamiokande:2021jaq}.
A new analysis method for efficient background reduction and an improved neutron tagging algorithm allowed to lower the energy threshold to 9.3~MeV.
The flux upper limit is shown in the figure by the exposure of 183~kton\,$\cdot$\,yr.
Furthermore, the combined analysis of all phases of SK was conducted, and the flux limit for electron antineutrinos above 17.3~MeV was set to 2.7~cm$^{-2}$~s$^{-1}$ with the exposure of 359~kton\,$\cdot$\,yr.
In this paper, a model-dependent search with 21 modern DSNB predictions was also conducted.
A 1.5$\sigma$ level of excess was observed over the background prediction across all the models, and the DSNB flux limits were between 2.5 and 2.8~cm$^{-2}$~s$^{-1}$.

Another water Cherenkov detector, Sudbury Neutrino Observatory (SNO), also reported some results of DNSB observations.
SNO was located in the Inco, Ltd. Creighton mine near Sudbury, Ontario, Canada, which used ultrapure heavy water.
By using heavy water, SNO enabled very unique neutrino observations, which could measure the charged current ($\nu_e + d \rightarrow p + p + e^-$) and neutral current ($\nu + d \rightarrow \nu + p + n$) interactions independently.
This led to conclusive evidence that neutrino oscillation also occurs in solar neutrinos~\cite{PhysRevLett.89.011301}.
As for the DSNB measurements, SNO showed the unique results that were the flux limit of electron neutrinos using the charged current interaction, instead of the electron antineutrinos which was rather high~\cite{PhysRevD.70.093014, Aharmim_2006}.
The result in the energy region between 22.9 and 36.9~MeV was 70~cm$^{-2}$~s$^{-1}$ with the exposure of 0.65 kton\,$\cdot$\,yr.

Recently, the results of DSNB observations by two liquid scintillator detectors were shown; one was KamLAND and the other was Borexino.
The KamLAND experiment uses 1~kton of ultrapure liquid scintillator located in Kamioka mine, Japan.
It started in 2002, including KamLAND-Zen, which set the balloon filled xenon-loaded liquid scintillator at the center in 2011 for the discovery of the neutrinoless double beta decay.
For the DSNB observations, the first results were published in 2012~\cite{Gando_2012}, followed by the results using more data and improved analysis methods in 2022~\cite{Abe_2022}.
In the latest result, the flux upper limit in the energy region between 8.3 and 30.8~MeV with the exposure of 6.72~kton\,$\cdot$\,yr was shown in the Figure~\ref{fig:flux_limit}.
It gave the most stringent flux limit below 13~MeV, though it was still an order of magnitude larger than DSNB theoretical expectations.
The Borexino experiment used 278~tons of ultrapure liquid scintillator located in the underground hall C of Gran Sasso, Italy.
Thanks to the exceptionally low level of radiopurity Borexino realized a low energy threshold.
It led to the first discovery of low-energy solar neutrinos such as $^7$Be, pep, pp, and CNO cycle.
For the DSNB search, neutrinos in a wide energy range from 1.8 to 16.8~MeV were searched~\cite{AGOSTINI2021102509}.
The flux upper limit with the exposure of 1.494~kton\,$\cdot$\,yr is shown in the Figure~\ref{fig:flux_limit}.
The flux limit below 8.3~MeV was set only by Borexino. 

\subsection{Backgrounds}
As described in the previous section, the DSNB has not yet been discovered.
It is limited by background events.
The DC method is effective for DSNB analysis to search for IBD events, and it has been recently become available for the water Cherenkov detectors as well as liquid scintillator detectors.
Nevertheless, there are still background events that can mimic IBD events.
At first, there are two major background sources: one is from spallation products induced by cosmic muons, and the other is from atmospheric neutrinos.
They are background events common to the two types of detectors, but there are also differences in the events observed in each type of detector.
In addition, reactor neutrinos can be background events in lower energy regions, and accidental coincidence backgrounds, which mimic DC events, should be considered in both types of detectors.
For liquid scintillator detectors, fast neutrons are also one of the backgrounds.
In this section, these backgrounds in the DSNB analysis are described.

\subsubsection{Spallation products}
Cosmic ray muons can penetrate detectors located deep underground, although the rate is drastically reduced relative to sea level.
These muons interact with the nuclei in the detector and produce various radioactive isotopes called `spallation products’.
The energies of the spallation products, $\beta$ and/or $\gamma$, are similar to those of positrons from DSNBs.
Most of these events can be removed using temporal and spatial correlations with muons.
Nevertheless, long-lived products are not easy to remove since it is quite difficult to identify their parent muon, and can eventually become background events in DSNB analysis.
In particular, $^9$Li, which undergoes beta decay (35\% of which emit one neutron) and has a higher production yield, is one of the remaining background events. 

\subsubsection{Atmospheric neutrinos}
When primary cosmic rays, which are mainly composed of protons, collide with nuclei in the Earth's atmosphere, several hadrons, such as pions and kaons, are generated.
Atmospheric neutrinos are generated when these hadrons decay.
The energy of atmospheric neutrinos are on the order of 100~MeV to PeV, peaking at several hundred MeV.
Particles generated by atmospheric neutrino interactions in the detector have a wide range of energies. If they have a reconstructed energy in the DSNB analysis region and neutrons are emitted together, will constitute a background event.
Both charged current (CC) and neutral current (NC) interactions cause such events.
Here, targets of neutrino interaction in the detector are nuclei (oxygen for water, carbon for liquid scintillator) and free protons.

As for the water Cherenkov detector, the most serious background is the neutral current quasi-elastic (NCQE) interaction, which knocks out a nucleon in oxygen nuclei.
If the knocked-out nucleon is a neutron and the excited nucleus generates gamma rays on deexcitation, it is indistinguishable from DSNB signal because the gamma-ray energy is similar to that of positron in IBD~\cite{PhysRevLett.108.052505}.
Accurate estimation of this background is not easy because of several uncertainties in the nuclear models.
In addition, other interactions, such as an NC interaction with one pion ($\nu + N \rightarrow \nu + N’ + \pi$) or a charged current quasi-elastic interaction (CCQE, $\nu + N \rightarrow l + N’$, where $N$ and $N’$ are nucleons and $l$ is a lepton), could be backgrounds. However, this is the case if neutrons are emitted together.

For liquid scintillator detectors, estimating the background NC interaction with carbon is also challenging due to its uncertainties.
The most dominant process is $\nu (\bar{\nu}) +^{12}C \rightarrow \nu (\bar{\nu}) + n + ^{11}C + \gamma$, but all possible interaction modes are considered in Ref.~\cite{Gando_2012}.
Compared to NC, the contribution of CC is not much.
In the case of CC, interaction with free proton is dominant over that with carbon nuclei because of their cross section~\cite{KIM2009330}.
Here, the muon antineutrino interaction is more likely to be the background in the DSNB analysis energy region than electron antineutrinos because a large fraction of the initial neutrino energy is spent to generate muon and the observed energy is shifted to low.

\subsubsection{Reactor neutrinos}
Reactor neutrinos are produced by the beta decay of neutron-rich fission fragments mainly from the following four isotopes: $^{235}$U, $^{238}$U, $^{239}$Pu, and $^{241}$Pu.
The flux is calculated from reactor operating data such as thermal power generation, fuel burn, shutdown, and fuel reload schedules.
The energy spectrum is calculated based on experimental results~\cite{PhysRevD.91.065002}.
Most reactor neutrinos have energies below 10~MeV, where the DSNB flux is still analyzed.

\subsubsection{Accidental coincidence}
In actual data acquisition, the prompt signal can be paired with a delayed signal that is misidentified due to radioactive event or noise, which is called `accidental coincidence'.
This background event is estimated for each experiment.

\subsubsection{Fast neutrons}
In liquid scintillator detector, fast neutrons produced by cosmic muons in the surrounding rock and water and introduced into the detector can also be background events.
Prompt events can be mimicked by the scattering of neutron on protons or carbon nuclei, followed by a delayed event when the neutron is thermalized and captured on a proton or carbon nucleus.

\subsection{Recent upgrade of Super-Kamiokande}
In the summer of 2020, Super-Kamiokande moved to a new phase with the start of SK-Gd, which drastically improves the detection sensitivity of DSNB by adding gadolinium.
As described above, neutron tagging in IBD is crucial for background reduction in water Cherenkov detector.
The original idea for this purpose in Super-Kamiokande was proposed in 2004~\cite{Beacom:2003nk}.
Gadolinium has the largest neutron capture cross section of all elements, giving, for example, a neutron tagging efficiency of 90\% at a mass concentration of only 0.1\%.
In addition, the delayed gamma-ray energy is about 8~MeV in total, which is high enough to be detected by Super-Kamiokande.
After years of research and development~\cite{MARTI2020163549} and a tank refurbishment work with the main purpose of repairing a water leak in 2018, 13~tons of gadolinium sulfate equivalent to 0.011\% mass concentration were loaded in 2020~\cite{ABE2022166248} as a first step, and this phase is called SK-VI.

The first result using about 1.5 years of data from SK-VI appeared in~\cite{Harada_2023}.
This analysis did not assume any spectral shape of the astrophysical electron antineutrino sources as well as the DSNB.
Thanks to a significant increase in neutron tagging efficiency $(35.6\pm 2.5\%)$ mainly with gadolinium capture and low misidentification probability ($\sim$10$^{-4}$), the IBD signal efficiency was twice that of the previous pure water phase.
The events observed in the searched neutrino energy range of 9.3~MeV to 31.3~MeV were consistent with the expected background, mainly due to atmospheric neutrinos and spallation products like $^9$Li.
Though no significant signal was discovered from this result, its detection sensitivity was comparable to that of the previous pure water phase, SK-IV, by only one-fifth of statistics.
The flux upper limit with 34.0 kton\,$\cdot$\,yr in SK-VI is shown in Figure~\ref{fig:flux_limit}.

In the summer of 2022, an additional 26~tons of gadolinium sulfate were loaded into the detector as a second step.
The loading work was successfully completed in about a month.
As a result, gadolinium equivalent to a mass concentration of approximately 0.03\% was introduced into SK, and this new experimental phase is called SK-VII.
The neutron tagging efficiency was confirmed to be 1.5 times higher than in SK-VI, which was consistent with expectation.
In addition to the improving neutron tagging efficiency, this phase will allow analysis with large statistics data assuming the spectral shape of each DSNB model.

\subsection{Experiments in the near future}
\subsubsection{JUNO}

JUNO (Jiangmen Underground Neutrino Observatory) is the next generation neutrino detector, which is located at Jiangmen in Guangdong province, China.
It is the largest liquid scintillator detector ever, with the primary goal of accurately measuring neutrino oscillations from reactor antineutrinos.
In addition, it will be a pioneering experiment to observe the DSNB signal for the first time together with SK-Gd in the next decade.
After many years of construction, it will be operational in 2023.
The detector is 20~kton of liquid scintilator with 17,612 20-inch and 25,600 3-inch photomultiplier tubes which are installed in the gaps between the 20-inch photomultiplier tubes to improve the energy resolution.
The signal detection method is the same, i.e, the DC technique with proton capture via IBD.

Backgrounds for the DSNB search to be considered are described above.
First, the $\bar{\nu}_e$ from reactor and atmospheric neutrinos are inevitable.
The energy region of reactor neutrinos is less than 10~MeV, while the energy region of atmospheric neutrinos increases with the neutrino energy.
Thus, the search for is DSNB is in the energy region in between.
Other backgrounds could be long-lived isotopes due to muon spallation, especially $^8$He and $^9$Li.
This is because the $\beta$-n decay of these isotopes is very similar to the IBD signal.
Nevertheless, the energy of the prompt signal due to the beta decay is relatively low, allowing this background to be ignored if an appropriate energy threshold is set.
External fast neutrons could also be a background in liquid scintillator detectors.
They are, however, expected to be removed by appropriate fiducial volume cuts.
As described above, estimation and reduction of background due to atmospheric neutrino NC interaction with carbon is challenging.
The experimental group has devised various methods to reduce this background and has demonstrated their effectiveness through simulation studies~\cite{Abusleme_2022}.

\subsubsection{Hyper-Kamiokande}
Hyper-Kamiokande is the next generation water Cherenkov detector located 8~km from the SK site.
It is now under construction and will start the operation in 2027.
It is based on a well-established technology with a fiducial volume of 187~ktons, which is 8.3 times larger than that of SK~\cite{protocollaboration2018hyperkamiokande}.
This huge volume of the detector allows DSNB observations with large statistics as well as other physics targets, thereby promising, for example, the observation of the DSNB spectrum.
The first phase uses pure water, thus, the energy threshold is as high as about 16~MeV due to the effect of spallation background.
Nevertheless, if a DSNB signal with sufficient statistics is observed in the higher energy region, for example, it is sensitive to the black-hole formation rate and can provide information on the history of star formation and its metallicity.
In the future, if the gadolinium loading is realized, the energy threshold for a DSNB search can be lowered to about 10~MeV.
This would make it possible to explore the history of supernova explosions back to an age with a redshift of about 1.
Thus, Hyper-Kamiokande has the potential to perform neutrino astronomy and cosmology with DSNB observations.

\section{Outlook and conclusions}

This is an exciting era for neutrino astrophysics.
The detection of extraterrestrial neutrinos from an astrophysical origin associated with stellar explosions is just around the corner. The diffuse supernova neutrino background (DSNB) provides valuable information about past core collapse, including the fraction of black-hole forming collapse and neutrino physics through flavor oscillations or other exotic interactions over cosmological baselines. While detecting the DSNB and studying supernova physics and neutrino properties is a challenging goal, recent progress in neutrino experiments makes this an achievable goal in the next decade. For instance, the recent upgrade of the Super-Kamiokande detector has significantly reduced background noise, bringing the first detection of DSNB within reach. This review has gathered essential information on DSNB physics and the latest updates from DSNB search experiments. In parallel, it is likely that we will see in the next decades the second detection of neutrinos from a core-collapse supernovae after SN~1987A that happened more than a quarter century ago. Besides the mere excitement of detection, it will bring plenty of physics and astrophysics to be studied using the rich multi-messenger observational data. The coming decades therefore hold great promise for detecting and studying supernova neutrinos, with potential exciting breakthroughs. 

\section*{Acknowledgments}
This work was supported by 
MEXT KAKENHI Grant Numbers, JP20H05850, JP20H05861 (SA), 
NSF Grant No.~AST-1908960 (NE, SH), No.~PHY-2209420 (NE, SH), NSF Grant No.~PHY-1914409 (SH), 
the U.S.~Department of Energy Office of Science under award number DE-SC0020262 (SH), JSPS KAKENHI Grant Number JP22K03630 (SH), 20H00162 (YK) and World Premier International Research Center Initiative (WPI), MEXT, Japan (SA, SH, and YK).


\bibliographystyle{utphys}
\bibliography{refs}


\end{document}